\documentclass{aa}
\usepackage{psfig}

\def\b{$\beta$}
\def\G{$\Gamma$}
\def\g{$\gamma$}
\def\s{$\sigma$}
\def\NH{$N_{\rm H}$ }
\def\be{\begin{equation}}
\def\ee{\end{equation}}
\newcommand{\ltsima} {$\; \buildrel < \over \sim \;$}
\newcommand{\gtsima} {$\; \buildrel > \over \sim \;$}
\newcommand{\lta} {\lower.5ex\hbox{\ltsima}}
\newcommand{\gta} {\lower.5ex\hbox{\gtsima}}

\def\approxlt{\mathrel{\hbox{\rlap{\lower.55ex \hbox {$\sim$}}
        \kern-.3em \raise.4ex \hbox{$<$}}}}
\def\approxgt{\mathrel{\hbox{\rlap{\lower.55ex \hbox {$\sim$}}
        \kern-.3em \raise.4ex \hbox{$>$}}}}

\begin{document}

  \thesaurus{03         
             (11.01.2;  
               11.06.2;  
               11.09.4;  
               11.14.1;  
               13.25.2)  
}

\title{ROSAT HRI observations of radio-loud AGN}  
\author{M. Gliozzi\inst{1}, W. Brinkmann\inst{1} 
\and S. A. Laurent-Muehleisen\inst{2}\thanks{Visiting Astronomer, Kitt Peak 
National Observatory, National Optical
Astronomy Observatories, which is operated by the Association of Universities
for Research in Astronomy, Inc. (AURA) under cooperative agreement with the
National Science Foundation}
, L.O. Takalo\inst{3}
\and A. Sillanp\"a\"a\inst{3}}
\offprints{M. Gliozzi,\\
 mgliozzi@xray.mpe.mpg.de} 
\institute{
Max-Planck-Institut f\"ur extraterrestrische Physik,
         Postfach 1603, D-85740 Garching, Germany
\and  University of California - Davis and the Institute for Geophysics and
Planetary Physics, Lawrence Livermore National Laboratory, Livermore, CA
94450 , USA
\and Tuorla Observatory, University of Turku, 215000 Piikki\"o,
Finland
}
\date{Received: ; accepted: }
   \maketitle
\markboth{M.~Gliozzi et al.: Final ROSAT HRI observations}
{M.~Gliozzi et al.: Final ROSAT HRI observations}

   \begin{abstract}
We present the results of three ROSAT HRI observations of AGN
expected to reside in clusters of galaxies. Although the 
exposures were truncated by the premature end of the
ROSAT mission,  valuable information
could be achieved which greatly improved upon the previous PSPC results.
 
For RGB 1745+398 we could separate the cluster emission from that
of the BL Lac and could confirm the cluster parameters obtained from
optical follow-up observations. In MRC 0625-536 the flux
from the central point source contributes less than 3\% to the total
X-ray flux and the eastern component of the dumbbell galaxy seems
to be the X-ray emitter. RXJ1234.6+2350 appears to be extended in
X-rays. The X-ray flux is centered on a quasar, but optical spectroscopy
indicates that the nearby radio galaxies reside in a previously unknown
cluster at redshift z$\sim$0.134.

\keywords{Galaxies: active -- 
Galaxies: fundamental parameters  
-- Galaxies: ISM -- Galaxies: nuclei -- X-rays: galaxies }
   \end{abstract}
%
\section{Introduction}

Compared to most previous X-ray missions the ROSAT satellite
provided a substantial increase in sensitivity and spatial 
resolution, together with the moderate spectral resolution of the
PSPC detector (Pfeffermann et al. 1986).
For even higher spatial resolution,
the ROSAT  High Resolution Imager 
(HRI, David et al. 1998) provided a spatial
resolution of about 5\arcsec~in the 0.1-2.4~keV band, although knowledge of
the spectral response is sacrificed.
    
In this paper we present the analysis of HRI observations of three
radio-loud AGN which are expected to reside in the
center of clusters of galaxies. The PSPC's spatial
resolution turned out to be  insufficient to resolve the AGN
contribution from the cluster emission.
The objects were proposed for the sixth cycle of ROSAT observations. In all
three cases, the envisaged observation time had not been completed 
when the satellite mission ended unexpectedly.
The accumulated limited signal-to-noise ratio of the data
is not sufficient to fully achieve the expected scientific goals
but it allows insights into the physical conditions of the systems
going beyond those obtained from the PSPC data.
 
In this paper we will discuss the  ROSAT HRI data of 
three different objects:
Sect. 1 deals with the BL Lacertae object RGB 1745+398
located near the center of a bright galaxy cluster. 
In Sect. 2 we discuss the
radio source MRC 0625-536  at the center of the Abell cluster
A3391 and, finally in Sect. 3, the region around the X-ray source 
RX J1234.6+2350 where we find three point like
radio objects spatially coincident with the ROSAT source.
  
 Spatial 
scales and luminosities are calculated assuming 
$H_0=50 {~\rm km ~s^{-1}~Mpc^{-1}}$, $q_0=0.5$ and $\Lambda=0$. 
Therefore 1 arcmin approximately
corresponds to 307 kpc for RGB 1745+398, to 86 kpc for MRC 0625-536, and to
187 kpc for RX J1234.6+2350.
 
\section {RGB 1745+398}
  
RGB 1745+398 was identified as a BL Lac object 
(Laurent-Muehleisen et al. 1998) in optical follow up
observations of a correlation between the ROSAT All Sky Survey (RASS)
and the Green Bank 5~GHz radio survey (Brinkmann et al. 1995).
The discovery of a blue arclike structure 8\arcsec ~ southeast
of the BL Lac indicates the presence of a cluster of galaxies
(Nilsson et al. 1999).
   
BL Lacs in clusters as gravitational lenses are relatively rare.
Because of their strong variability they are 
 excellent candidates for measuring $H_0$ via gravitational time delay.
Interestingly, if all current lens candidates were confirmed, the incidence of
lensing would be much higher than in comparable, low z,  surveys of
quasars (Scarpa et al. 1999). 

The location of this nearby BL Lac object (z = 0.267) in the
 center of a cluster suggests that some portion of the X-ray emission 
 may be coming from the intra-cluster gas.
 One of the aims of this work is to check this hypothesis
and, eventually, to derive the gas and the total mass of the cluster 
from the X-ray data, independent of the optical results.

RGB 1745+398 was observed with the ROSAT HRI
between March 4 and 7,  1998  with an  effective exposure of 42.6 ksec.
 Only photons in the pulse height 
channels 2 to 8 are used in the spatial analysis in order to increase the
signal to noise ratio, as channels 9 to 15 mostly contain instrumental
background. 

A contour plot of the total X-ray emission overlaid on the optical image
is shown in Fig. 1. The photons were binned in two arcsec pixels and
subsequently smoothed with a Gaussian with \s~= 6 arcsec.

\begin{figure}
\psfig{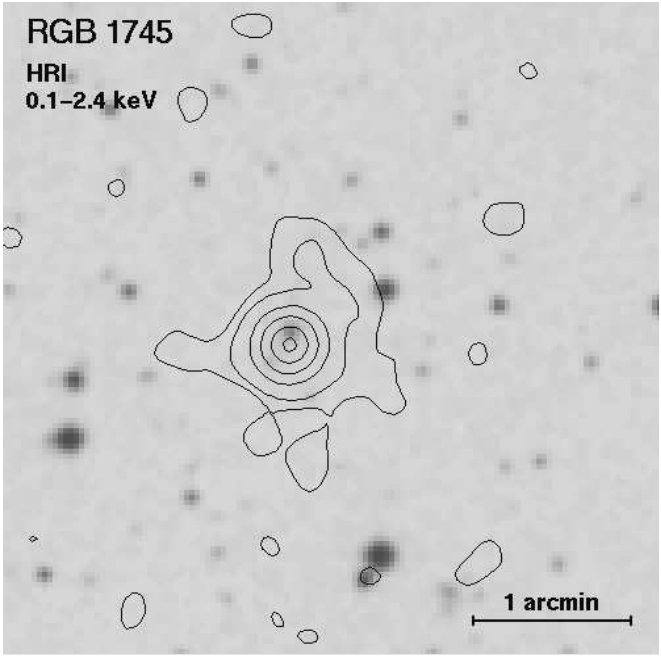}
\caption {Contour plot for the X-ray emission of RGB 1745+398 overlaid
on the optical image of a region $4\farcm16 \times 4\farcm16 $.
The contours correspond to 2.5, 5, 20, 50, 100, and 150 \s~ above background.}
\end{figure}

The X-ray emission appears to be extended and centered near to the arc-like
optical structure (Nilsson et al. 1999).
The angular separation of  about 4\arcsec ~ from the position of
the BL Lac is within the intrinsic resolution of the HRI of $\sim$ 5\arcsec
(Harris et al. 1998).  
 To determine the physical properties
of the extended X-ray emission we fitted a  \b-model (e.g. Cavaliere \&  
Fusco-Femiano 1976, Gorenstein et al. 1978) of the form 
\be
S(r)=S_0\left(1+{r^2\over r_c^2}\right)^{-3\beta+1/2}
\ee
to the surface brightness profile of the HRI source; the result is shown
in Fig. 2.
\begin{figure}
\psfig{figure=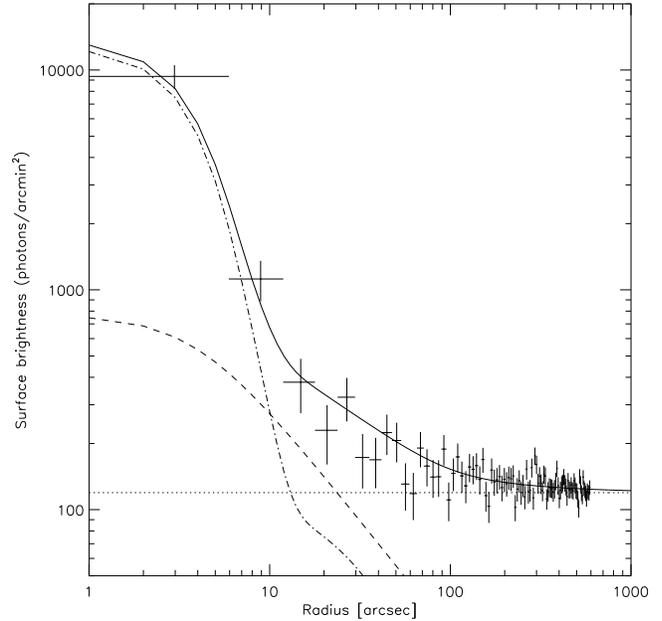,width=8.7cm}
\caption{The surface brightness profile for RGB 1745+398. The best fit
\b-model is indicated by the long-dashed line, whereas the background level
and the PSF model for the central point source are given by the dotted and
the dash-dotted lines, respectively.}
\end{figure}

The best fit values ($\chi^2_{\rm red}=1.04$)
for the \b-model are
$S_0=771 \pm 250 {\rm~cts/arcmin^2}$, $\beta=0.34 \pm 0.02$, $r_{\rm c}=3.9 \pm 1.7
{~\rm arcsec~} (20.2 \pm 8.7 {~\rm kpc})$. It must be noted that although the
value of \b~ is somewhat low with respect to the typical value for clusters
($\sim 0.6$), the core radius $r_c$ is consistent with that obtained 
from the optical data (Nilsson et al. 1999). 
To account for the BL Lac emission
we included a Point-Spread-Function (PSF) model in the fit and convolved the
original PSF model with an additional Gaussian to allow for the known smearing
of the PSF by residual wobble motion, which varies between different
observations (Morse 1994). The normalization turns out to be $15690 \pm
300 {\rm~cts/arcmin^2}$, and the additional
 \s$_{\rm wobble}$ is 1.5$\pm 1$ arcsec. By 
integrating the two profiles we derive that the BL Lac object and the extended
region give nearly the same contribution to the total X-ray emission, with
a slight excess from RGB 1745+398 (52\% vs. 48\%). 
Given that the HRI lacks spectral response,
we calculated the 0.1-2.4 keV flux for the 
extended emission assuming Galactic absorption and a thermal
Bremsstrahlung spectrum with $kT$ ranging between 1~keV and 10 keV.
For the point source we assume a power law spectrum with a photon
index \G~ranging between 1.5 and 2.5.  With these
assumptions the flux ranges of the extended and point-like source are 
$f_{\rm X,ext}\simeq (3.7-3.9)\times 10^{-13} {~\rm erg~cm^{-2}s^{-1}}$ and
$f_{\rm X,point}\simeq (4.1-6.4)\times 10^{-13} {~\rm erg~cm^{-2}s^{-1}}$,
respectively.
  
The corresponding luminosities are shown in Fig.3.
It can be seen that, even for the generally steep power law spectra 
found in the ROSAT band  for BL Lacs (\G \gta 2.2; Brinkmann et al. 1996),
 RGB~1745+398
has a rather low X-ray luminosity and, according to its 
radio - to - X-ray index  $\alpha_{\rm rx} \sim 0.8$, it belongs more
to the class of Low energy peaked BL Lacs (Giommi \& Padovani 1994).

\begin{figure}
\psfig{figure=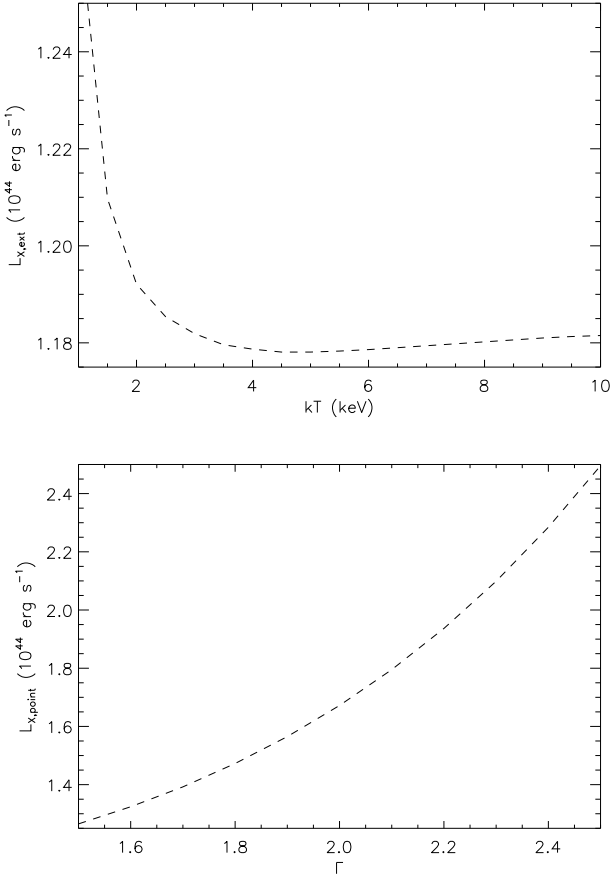,width=8.7 cm}
\caption{ X-ray luminosities of the extended thermal component (upper panel) and 
the point-source (lower panel), respectively as function of their spectral
parameters.}
\end{figure}

The physical parameters of the cluster, the central density, the gas mass,
and the total mass can be obtained by de-projecting the surface brightness
profile. With this method one can derive the corresponding density profile:
\be
n(r) = n_0 \left(1+{r^2\over r_c^2}\right)^{-{3\beta\over 2}}
\ee 
where the central density $n_0$ is given by (e.g. Henry et al. 1993): 
\begin{eqnarray}
n_0 & = & 1.2\times 10^{12} {~\rm cm^{-3}}
\left[{I_0\over r_c(kT)^{1/2}}\right]^{1/2}\times \nonumber \\
    & ~ &\left\{ {\Gamma(3\beta-1/2)\over\Gamma(3\beta)}
\left[\gamma\left(0.7,{E_2\over kT}\right) - 
\gamma\left(0.7,{E_1\over kT}\right)\right]\right\}^{-1/2} 
\end{eqnarray}     
$I_0$ is the central intensity $S_0$ converted to {$~\rm  erg~cm^{-2}s^{-1}
sr^{-1}$}, $r_{\rm c}$ the core radius in cm and $kT$ is given in keV.
\G~ and \g~ are the complete and incomplete Gamma functions respectively.
$E_1$ and $E_2$ denote the observed energy range, i.e. 0.1 keV and 2.4 keV
in case of the ROSAT HRI observations.
 For the conversion from $S_0$ to $I_0$
we assumed a thermal bremsstrahlung spectrum with $kT=4$ keV and Galactic
absorption. Using the best-fit parameters from the \b-model, we obtain
$n_0=1.04\times 10^{-2} {~\rm cm^{-3}}$. The total gas mass within a given
radius can be calculated by integrating Eq. (2). Assuming a radius
of 8 arcsec ($\sim$ 41 kpc), corresponding to the Einstein radius 
of the cluster
(Nilsson et al. 1999), the total gas mass ranges from  $1.2\times 10^{10}
{\rm M_\odot}$ ($kT=10$ keV) to
$3.4\times 10^{11} {\rm M_\odot}$ ($kT=1$ keV).

Assuming hydrostatic equilibrium for the intra-cluster medium of RGB 1745+398
and spherical symmetry, we use the temperature and gas density profiles to
derive the total gravitating mass as a function of the radius:
\be
M_{\rm tot}= -{kT_g(r)r\over G m_H\mu}\left({r\over T_g(r)}{dT_g(r)\over dr} +
{r\over\rho (r)}{d\rho(r)\over dr}\right)
\ee

As discussed by Neumann \& B\"oringer (1995), deviations from the hydrostatic
equilibrium as well as moderate ellipticities do not have strong effects
 on the mass determination. With the additional assumption of an isothermal
intra-cluster medium Eq. (4) reduces to
\be
M_{\rm tot}={3\beta kT_g(r)r\over G m_H\mu r_c^2}{r^3\over 1+r^2/r_c^2}
\ee

For the total gravitating mass within 41~ kpc, we get $M_{\rm tot}=1.6
\times 10^{12} - 
 1.6\times 10^{13} {\rm M_\odot}$ for the cases  kT=1~keV  or 10~keV,
respectively. These values are consistent with those derived  from the
optical data ($1.3 \times 10^{13} {\rm M_\odot}$).
As a consequence the gas mass fraction at this radius ranges between 0.1\%
and 2.1 \% of the total cluster mass. 
 
The low total mass indicates that we see a poor cluster or a group of galaxies
and the obtained soft X-ray luminosity as well as
the low gas mass fraction are consistent with this picture (Reiprich
\& B\"oehringer 1999).
These results are in accordance with the claim that BL Lac objects are 
frequently found in groups of galaxies or in relatively poor clusters 
 with richness class $<$ 0 (Wurtz  et al. 1996).  

\section{MRC 0625-536}
Previous ASCA/ROSAT observations of the nearby (z= 0.0539) radio galaxy 
MRC 0625-536 (Otani et al. 1998) revealed 
that the X-ray emission from the system 
is largely dominated by the surrounding cluster Abell 3391, 
with only a small contribution 
($\leq 10\%$) due to the galaxy. With the superior spatial resolution of the
HRI, we tried to constrain the real X-ray luminosity fraction 
attributable to the dumbbell galaxy ESO~161-IG~007, the optical counterpart
of the radio source,  and to compare the X-ray brightness distribution to the optical 
and radio images to see which of the two galaxy components is active.

MRC 0625-536 was observed with the ROSAT HRI during the period 15-25 March
1998. The total exposure was 15.2 ksec. Again, only photons in 
the pulse height channels 2 to 8
were used in order to increase the signal to noise ratio.

A contour plot of the X-ray emission  overlaid on the optical image 
is shown in Fig. 4.
The photons were binned in four arcsec pixels and subsequently smoothed with
a Gaussian with \s = 12 arcsec.
>From the figure  it can be seen that the X-ray emission is clearly 
concentrated on the eastern galaxy, with a more extended structure near the
western galaxy, and there seem to be other point-like sources inside the cluster
region. In particular, there is a relatively strong X-ray source  at 
 RA(2000)=$6^h26^m36\fs7$, DEC(2000)=$-53^o40\arcmin39\farcs18$, detected
with the maximum likelihood technique, located $\approx 2.5$ arcmin 
north-east from the dumbbell galaxy, whose X-ray flux is comparable to that
of MRC 0625-536. This object is classified as 'Stellar' with a magnitude of
 $m_B = 15.3$ in the 
COSMOS digitized optical survey of the southern sky, operated by the ROE/NRL.
A search for a known object at this position
in the NASA/IPAC Extragalactic Data Base (NED) and in SIMBAD
gives negative results.

\begin{figure}
\psfig{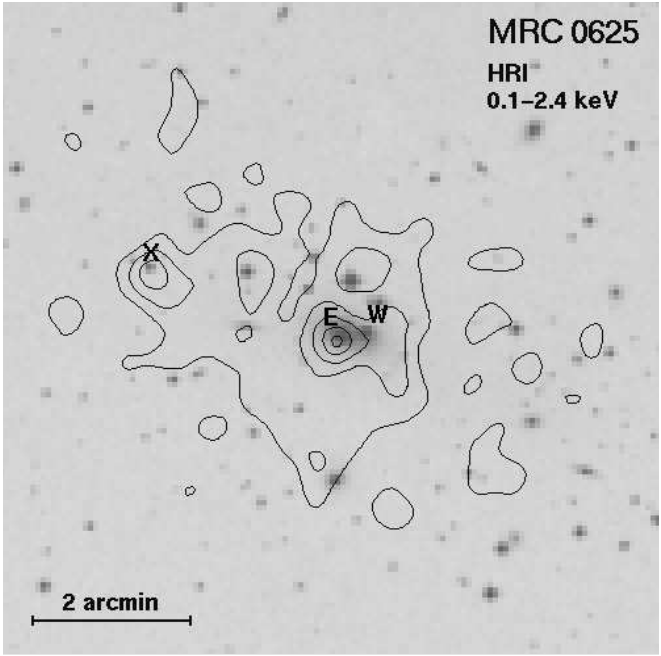}
\caption {Contour plot for the X-ray emission of MRC 0625-536 overlaid
on the optical image of a region $8\farcm33 \times 8\farcm33 $.
The contours correspond to 4, 7, 10, 13 and 15 \s~ above background. 
The eastern and western components of the dumbbell galaxy are labeled by
E and W respectively, while X marks the position of an additional unidentified
X-ray source.}
\end{figure}

To check the consistency of the HRI data with
previous ASCA/ROSAT results, we fitted 
the surface brightness profile with the \b-model described above.
The $\chi^2$  fit results ($\beta=0.56\pm 0.05,~r_c=218\pm 20 {~\rm kpc}$) 
are in perfect agreement with those obtained by Otani et al.
(1998, $\beta=0.56\pm 0.04,~r_c=241\pm 43 {~\rm kpc}$). 
In order to determine the X-ray fluxes and luminosities associated with both
the point-like and the extended sources we used the best fit results
obtained in the joint ROSAT and ASCA spectral analysis. 
In particular, we took 
$N_{\rm H}=4.6 \times 10^{20} {\rm~cm^{-2}}$ and a thermal model with 
$kT=5.7$ keV for the extended emission, while a power law model with
 photon index
$\Gamma=1.67$ was assumed for the dumbbell galaxy. 
In a circular region with radius
of  6~arcmin, the 0.1-2.4 keV flux associated with the extended region 
is $f_X\simeq 1.2\times 10^{-11} {~\rm erg~cm^{-2}s^{-1}}$, while for the
dumbbell galaxy we obtain $f_X\simeq 3.2\times 10^{-13} 
{~\rm erg~cm^{-2}s^{-1}}$. The corresponding luminosities are
$L_{\rm X,ext}\simeq 1.5\times 10^{44} {~\rm erg~s^{-1}}$ and 
$L_{X}\simeq 4.04\times 10^{42} {~\rm erg~s^{-1}}$. 
Therefore, the contribution from MRC 0625-536 
 to the total X-ray emission from  the system  is only about 3\%.

\begin{figure}
\hspace*{4.2cm}
\makebox[0cm]{\psfig{file=H1719f5a.ps,bbllx=89pt,bblly=303pt,bburx=507pt,%
bbury=714pt,angle=0,width=8.3cm,clip=0}}%
\makebox[0cm]{\psfig{file=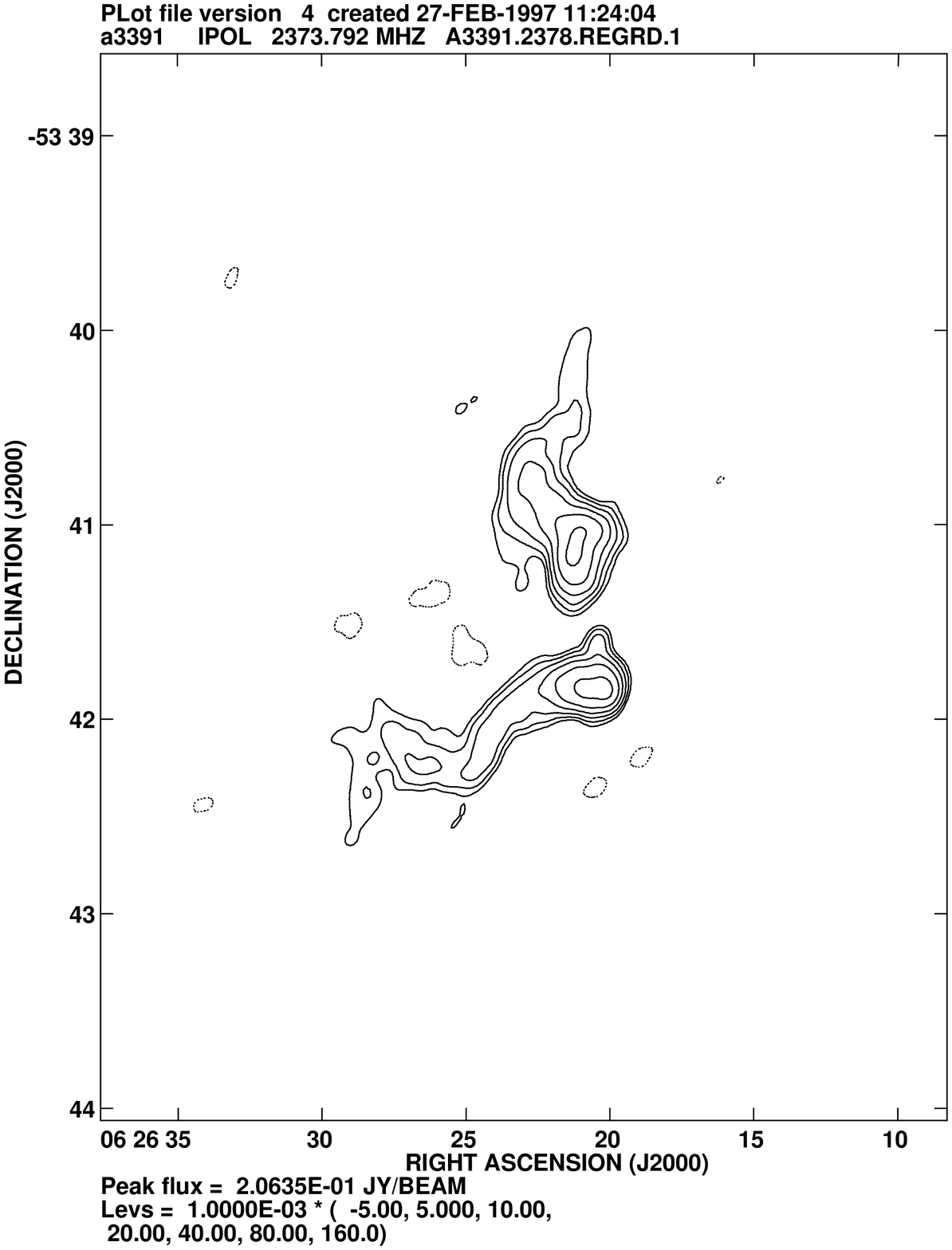,bbllx=125pt,bblly=180pt,bburx=565pt,%
bbury=613pt,angle=0,width=8.3cm,clip=0}}
\vskip 0.2cm
\caption{13 cm ATCA radio contours overlaid on the HRI image of MRC 0625-536.
The peak flux density is 206 mJy/beam, with contour levels at 5, 10, 20, 40, 
80 and 160 mJy/beam. The coordinate grid has spacing of 1 arcmin in declination
and 10 s in right ascension.}
\end{figure}

In Fig. 5 we show an overlay of the ATCA 13 cm  (2.3 GHz) radio contour map 
on a grey scale image of the central $\sim 4\times 4$ arcmin of the X-ray 
emission. 
The radio emission clearly originates from the eastern galaxy, where
the X-ray brightness shows a peak as well. Both jets are deflected, in 
particular the northern jet seems to be bent by a region of enhanced X-ray
emission. However, the limited signal to noise ratio of the HRI data does not
allow to improve previous conclusions on the large scale structures 
and the  complex interaction of the radio jet with the surrounding 
 cluster gas, obtained
from the PSPC observations (Otani et al. 1998).

\section{RX J1234.6+2350}

\begin{table*} 
\caption{Properties of the radio galaxies surrounding RX J1234.6+2350}
\begin{center}
\begin{tabular}{lllccc}
\hline
\noalign{\smallskip}
Source & RA(2000.0) & DEC  & $F_{\rm int}$ (1.4 GHz) & $F_{\rm X}$ 
(0.1-2.4 keV) &  B magnitude\\
\noalign{\smallskip}
       &  h~~~m~~~s   & $^o$~~~'~~~"      &  mJy          
       & ${\rm erg~cm^{-2}s^{-1}}$ &  \\
\hline
\hline
\noalign{\smallskip}
\noalign{\smallskip}
RX J1234.6+2350 & 12 34 39.1 & 23 50 05.0 &  &
 $2.1 \times 10^{-12}$ & 17.2\\
\noalign{\smallskip}
FIRST J123438.6+235013 & 12 34 38.63 & 23 50 13.0 & 4.54 &    & 19.3 \\
\noalign{\smallskip}
FIRST J123437.1+235016 & 12 34 37.13 & 23 50 16.1 & 3.48 &    & 19.1 \\
\noalign{\smallskip}
FIRST J123438.2+234947 & 12 34 38.27 & 23 49 47.6 & 8.22 &    & 19.1 \\
\hline
\end{tabular}
\end{center}
\end{table*}

This object was first seen in the correlation between the ROSAT 
All-Sky Survey (RASS) and the Green Bank 5~GHz survey 
(Brinkmann et al. 1997) and optically identified by Bade et al. (1998)
as a quasar although no redshift is given.  
In the FIRST 1.4~GHz VLA Survey (Faint Images of the Radio Sky at Twenty 
centimeters; Becker et al. 1995) three  radio sources, each with its own
optical counterpart, were found within the RASS position
error circle of RX J1234.6+2350 (Brinkmann et al. 1999). 
This constellation, an X-ray source surrounded by several faint radio sources,
occurs relatively often in the correlation of the FIRST and RASS catalogs.
It appears possible that  in these cases the X-ray emission originates from a 
distant cluster of galaxies while the sensitive radio observations 
trace individual galaxies in the cluster.  

RX J1234.6+2350 was observed with the ROSAT HRI between June 30 and July 1,
 1998 when the satellite was in a very unstable status.
Due to the short effective exposure (4.7 ksec) the data are only of
limited quality.  An  attempt to fit the X-ray
 emission with the \b-model failed, probably because of the low
photon statistics. Nevertheless the
X-ray contours (obtained after binning the photons in two arcsec pixels and
subsequent smoothing with a Gaussian with \s~= 6 arcsec)
overlaid on the optical image (Fig. 6) show some interesting
features. 

\begin{figure}
\psfig{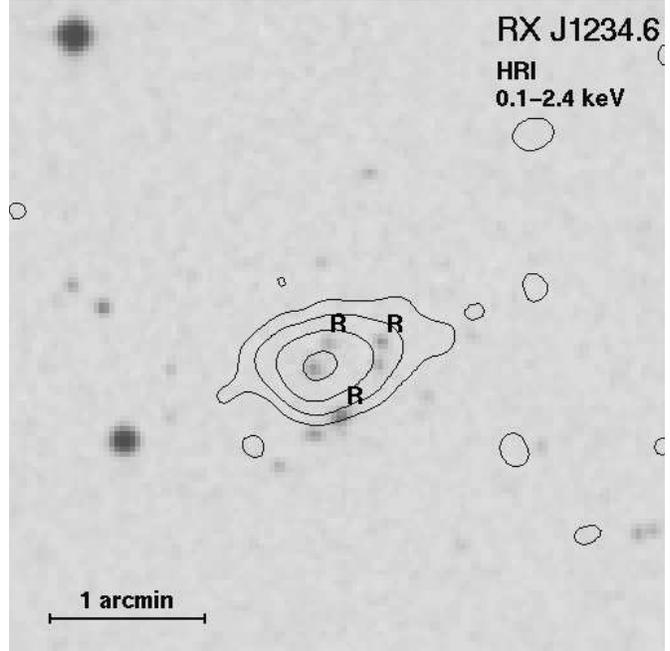}
\caption {Contour plot for the X-ray emission of RX J1234.6+2350 
(FIRST 1235+2350) overlaid
on the optical image of a region $4\farcm16 \times 4\farcm16 $.
The contours correspond to 3, 10, 30 and 80 \s~ above 
background. The letters R label the three radio sources falling
inside the RASS position error circle of RX J1234.6+2350.}
\end{figure}

The $1'\times 1'$ region surrounding the X-ray source is crowded 
with the quasar RX J1234.6+2350 and three, optically resolved, radio galaxies, 
whose coordinates and fluxes are given in Table 1.
  
We obtained new low dispersion optical spectra on June 7, 1997, of the three
radio sources in the field at the Kitt Peak National Observatory's 2.1-m
telescope using the GoldCam spectrograph.  Spectra were taken through a
1.5$\arcsec$ slit, resulting in a resolution of 4\,\AA.
Reduction proceeded in the standard manner using the IRAF (V2.11) analysis
package.  Wavelength calibration was carried out using comparison lamps 
generally taken at the beginning of the night.  The spectra had to be smoothed
to determine (by cross-correlation with a galaxy template) a redshift, which 
is estimated to be accurate to $\pm$0.002.  Fig. 7 shows the
spectra of the three radio sources given in Table 1.

\begin{figure}
\psfig{figure=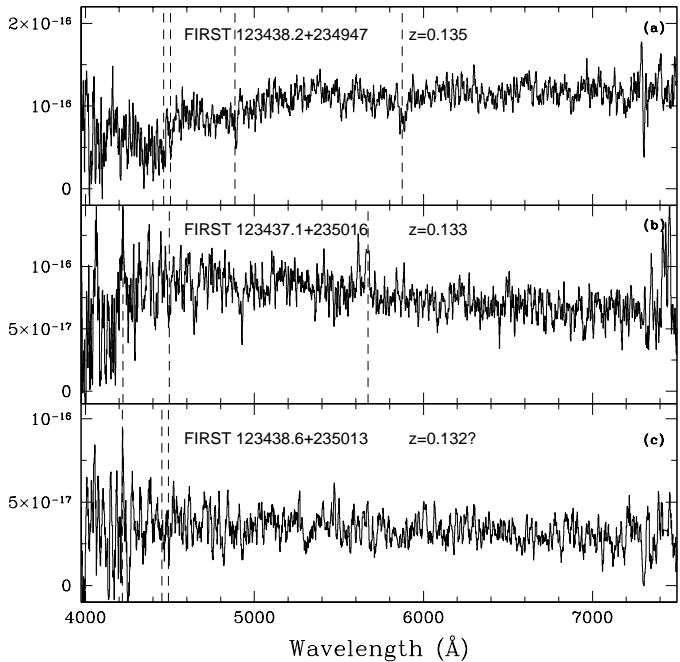,width=9.0truecm,angle=0,%
bbllx=18pt,bblly=144pt,bburx=567pt,bbury=693pt,clip=}
\caption{New spectra for radio sources in the field of RXJ1234.6+2350.  The
spectra (a) -- (c) have been smoothed by 5, 5 and 7 pixels, respectively.
Vertical dotted lines correspond to features used to confirm the redshifts
determined by cross-correlation with galaxy templates.  From shortest
wavelengths to longest: (a) Ca II H \& K, G band and MgIb, (b) [OII] (in
emission), Ca II K and [OIII] (in emission), (c) [OII] (in emission), Ca II H
\& K.}
\end{figure}
Although the spectra are of relatively low signal to noise ratio, the redshifts 
of FIRST 1234.2+234947 and FIRST 123437.1+235016 are well determined and 
consistent with being the same, given our errors.  We also believe that
the redshift of FIRST 123438.6+235013 is consistent with belonging to this
cluster, but a better spectrum is required to confirm this.
  
The X-ray emission is extended and the radial intensity distribution
is not consistent with the HRI's  point spread function.
However, the satellite's attitude
solution during these late phase HRI observations was not free of
problems and there is no other strong source in the field of view
to check  the effective point spread function.
In the ROSAT All Sky Survey the source appears extended as well
with a size of $\approx$ 1\arcmin ~and a peculiar morphology which
could indicate the presence of multiple, unresolved  sources.
But the limited photon statistics and the PSPC's spatial resolution
do not allow us to make any definite conclusions.

The X-ray flux in the ROSAT energy band (0.1-2.4 keV), calculated from
a circular region with radius $\approx 1$ arcmin centered on RX J1234.6+2350,
assuming a power law spectrum with photon index \G~=2.0 and Galactic absorption
towards the source of \NH=1.29 {$\rm \times 10^{20} cm^{-2}$}
(Dickey \& Lockman 1990),
is $f_X\simeq 2.1\times 10^{-12} {~\rm erg~cm^{-2}s^{-1}}$. 
 Assuming that the extended X-ray emission originates from a cluster and that
the latter is at the redshift of the radio galaxies
(z$\simeq 0.134$), the X-ray luminosity deduced from the HRI 
data is  $L_X=1.7\times 10^{44}{~\rm erg~s^{-1}}$.
  
The X-ray flux obtained from the HRI observation is nearly
a factor four lower than the flux obtained from the RASS count rate ($0.183\pm
0.022$ counts/s, corresponding to a flux of $8.5\times 10^{-12} {~\rm erg~
cm^{-2}s^{-1}}$). This implies that either the source is strongly variable
or that the spectral shape is completely different, leading to 
greatly differing count rates in the two detectors. However,
the survey data are consistent with a power law with slope 
$\Gamma = 1.94\pm0.18$ (Brinkmann et al. 1997).
  
The  HRI X-ray contours are centered on the quasar RX J1234.6+2350 
which might be the dominant contributor to the X-ray flux.
Its optical identification and B magnitude were obtained from the
Hamburg/RASS Catalogue (Bade et al. 1998).
Taking the optical flux at 2500\AA ~and  assuming an upper limit
$f_{\rm lim}\leq 1$ mJy
to the  1.4~GHz flux density at the quasar position,
which is converted to a 5~GHz flux density by assuming a power law
spectrum with $\alpha_{\rm r} = 0.5$, we obtain for the radio-loudness of
the quasar an upper limit of 
$R_{\rm l} = \log(F_{\rm r}/F_{\rm opt}) < -0.15$, 
which means that  RX J1234.6+2350 is a radio-quiet quasar.

\section{Conclusions}
 
Previous PSPC observations indicated that the objects observed with
the HRI are AGN residing in clusters of galaxies. As the
spatial resolution of the PSPC was insufficient to separate the
cluster emission from that of the AGN  we conducted ROSAT HRI
observations to study the spatial distribution of the 
X-ray emission in more detail.
 
In two of the three cases we can determine the fractional flux contribution
of the AGN to the X-ray flux although the HRI 
exposures were truncated by the premature end of the
ROSAT mission.
For RGB 1745+398 we could separate the cluster emission from that
of the BL Lac and could confirm the cluster parameters obtained from
the optical follow-up observations. The BL Lac object contributes 
about 48\% to the total soft X-ray emission. 
 In MRC 0625-536 the flux
from the central point source amounts to less than 3\% of the total
X-ray flux and it appears that the eastern component of the dumbbell galaxy
ESO~161-IG~007  is the X-ray emitter.
 
 RXJ1234.6+2350 seems to be extended 
with the X-ray flux  centered on a quasar. The optical spectra indicate
that the three nearby radio galaxies are members of a previously 
unknown cluster,
but no further data are available from this unusual system. The limited quality
of the X-ray data does not allow a decomposition of the quasar flux from the
extended emission component.
The changes of the X-ray flux between the ROSAT survey and the pointed 
observation strongly favor the quasar as the main contributor to the 
X-ray emission although for radio-quiet quasars flux changes by a factor
of four are rare (Yuan et al. 1998). 

X-ray observations with high spatial resolution are a 
powerful method to study these interesting kind of systems, AGN 
in the centers of  clusters, which show indications for
a direct interaction of the AGN activity with the surrounding matter.

\begin{acknowledgements}
The ROSAT project is supported by the Bundesministerium f\"ur
Bildung, Wissenschaft, Forschung und Technologie (BMBF) and
the Max-Planck-Gesellschaft.  We thank our colleagues from the
ROSAT group for their support. We wish to thank Robert Becker for taking the 
Kitt Peak spectra of the radio sources in the RX J1234.6+2350 field.
MG  acknowledges partial support from
the European Commission under contract number ERBFMRX-CT98-0195 
(TMR network ``Accretion onto black holes, compact stars and protostars").
SALM acknowledges partial support from
the U.S.\ Department of Energy at the Lawrence Livermore National Laboratory
under contract W-7405-ENG-48i and from the NSF (grant AST-98-02791).
This research has made use of the NASA/IPAC Extragalactic Data Base
(NED) which is operated by the Jet Propulsion Laboratory, California
Institute of Technology, under contract with the National Aeronautics
and Space Administration.
\end{acknowledgements}

\end{document}